# Classification of Pressure Gradient of Human Common Carotid Artery and Ascending Aorta on the Basis of Age and Gender

Renu Saini
Student
Electrical, Electronics and Communication Engineering Department,
The Northcap University, Gurgaon,122017
Haryana, India

Sharda Vashisth, PhD
Associate Professor
Electrical, Electronics and Communication Engineering Department,
The Northcap University, Gurgaon,122017
Haryana, India

Ruchika Bhatia
Student
Electrical, Electronics and Communication Engineering Department,
The Northcap University, Gurgaon,122017
Haryana, India

## ABSTRACT
The current work is done to see which artery has more chance of having cardiovascular diseases by measuring value of pressure gradient in the common carotid artery (CCA) and ascending aorta according to age and gender. Pressure gradient is determined in the CCA and ascending aorta of presumed healthy volunteers, having age between 10 and 60 years. A real 2D model of both aorta and common carotid artery is constructed for different age groups using computational fluid dynamics (CFD). Pressure gradient of both the arteries are calculated and compared for different age groups and gender. It is found that with increase in diameter of common carotid artery and ascending aorta with advancing age pressure gradient decreases. The value of pressure gradient of aorta is found less than common carotid artery in both cases of age and gender.

## General Terms
Ageing, Blood flow, Common Carotid artery, aorta, Pressure gradient, CFD

## Keywords
Ageing, Blood flow, Common Carotid artery, aorta, Pressure gradient

## 1. INTRODUCTION
A lot of human death is caused due to cardiovascular diseases which are related to the heart or blood vessels (arteries and veins). The reason behind increase in risk of cardiovascular diseases is pressure gradient. In order for blood to flow through a vessel or across a heart valve there must be a force driving the blood. This force is the difference in blood pressure (i.e., pressure gradient) across the vessel length or across the value which is given by Eq. 1

$F = \Delta P/R$ ......... (1)

Where F is the force driving the blood, $\Delta P$ is pressure gradient and R is resistant to flow of blood. So pressure gradient is directly proportional to force driving the blood. A normal valve especially normal large artery has small resistance to flow and therefore the pressure gradient across the valve is very small. But in case of vascular artery pressure gradient is more due to increased resistance to flow.

The early research was conducted using high resolution multigate pulsed Doppler system on the men carotid artery which shows changes in wall property of carotid artery according to age [11] and as age increases diameter and stiffening of the vessel wall also increases [8,11]. In further investigations the level of blood pressure was calculated in the common carotid artery (CCA) as function of age and gender along with interactions between diameter and storage capacity using high resolution ultrasound system [13]. Piezoelectric sensor was also used to see age associated developments in CCA [6, 7, 10]. The pulsed Doppler system, high resolution ultrasound systems or piezoelectric sensor systems were limited due to their heavy investments in equipment.

In recent years, due to its fast and economical feature, computational fluid dynamics (CFD) has become popular research tool for modelling flow behaviour within arteries. Investigators perform fluid and structural responses to pulsatile Non-Newtonian blood flow through stenosed artery using ANSYS. They assume blood behaviour as per Carreau Non-Newtonian model to estimate wall shear stress [14].

Pressure gradient is also assessed in aorta. In the early research Magnetic resonance imaging (MRI) method were used to collect subject specific geometry and flow rates in a human aorta [5]. Further investigation show geometric changes of the aortic arch according to age [1].

Since CFD is more economical and fast method recently it is used to measure wall shear stress (WSS) of healthy human aorta and age-related changes in aortic 3D velocity [9].

The objective of the present investigation is to study the age associated developments in pressure gradient in common carotid artery and ascending aorta of normal humans having age between 10 and 60 years and then comparing the pressure gradient of both arteries to see which artery has more chance of having cardiovascular diseases. Here CCA is compared with aorta because at aortic arch aorta bifurcates into three arteries i.e.; brachiocephalic artery, carotid artery and left subclavian artery. The oxygenated fresh blood from the heart via left ventricle enters the ascending aorta and then it is directed toward the aortic arch and comes out through descending aorta. From aortic arch blood is directed to CCA which direct blood to face, brain. Therefore classification of pressure gradient in aorta is also done on the basis of age and gender.

## 2. METHOD

Most commonly used analysis tool nowadays is Computational Fluid Dynamics (CFD). This tool is used to model behaviour of flow within arteries. This tool is fast and economical method to analyse and fix problem as it used computers to perform simulation instead of heavy and costly equipment. In this simulation is done using some boundary conditions.

For different age group and gender as shown in table 1, 2 and 3 a real 2D model of both aorta and common carotid artery is constructed [9, 13] using CFD as shown in Fig. 1 and 2. The practical block diagram for simulation is shown in Fig. 3.

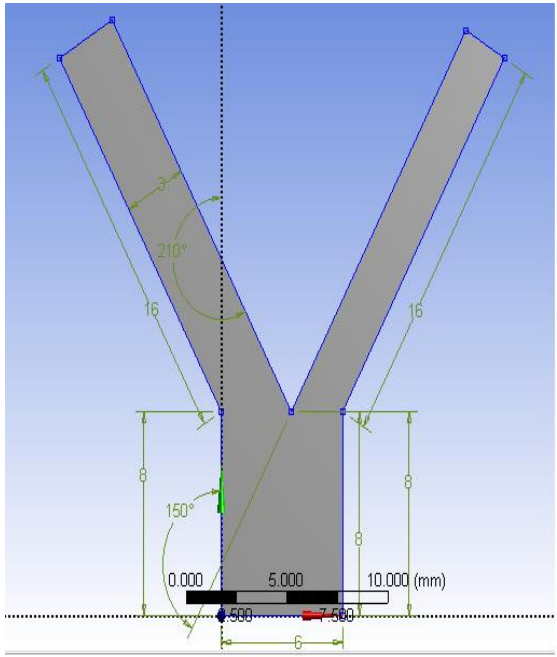

**Fig. 1: Geometry of common carotid artery**

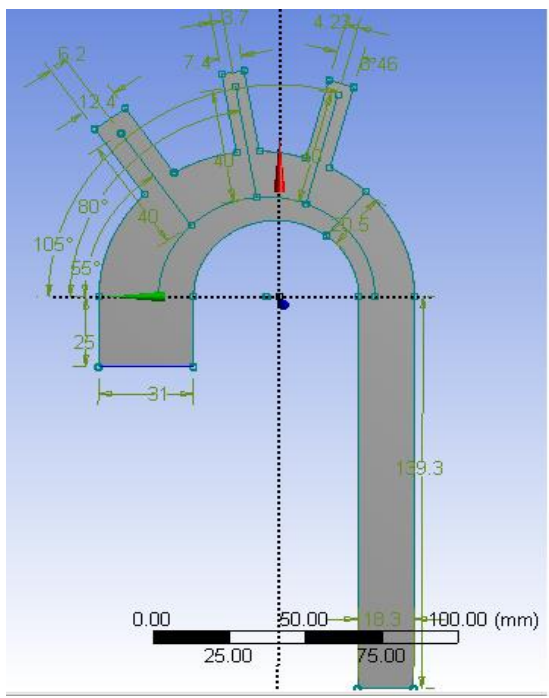

**Fig. 2: Geometry of aorta**

**Table 1: Diameter of Common carotid artery with age for female group**

| SNo. | Age Group (Years) | Diameter (mm) |
|---|---|---|
| 1 | (10-20) Years | 5.9 |
| 2 | (30-40) Years | 6.2 |
| 3 | (50-60) Years | 6.3 |

**Table 2: Diameter of Common carotid artery with age for male group**

| SNo. | Age Group (Years) | Diameter (mm) |
|---|---|---|
| 1 | (10-20) Years | 6.0 |
| 2 | (30-40) Years | 6.5 |
| 3 | (50-60) Years | 6.4 |

**Table 3: Diameter of Ascending Aorta with age**

| SNo. | Age Group (Years) | Diameter (mm) |
|---|---|---|
| 1. | (20-30) Years | 28 |
| 2. | (30-40) Years | 31 |
| 3. | (40-50) Years | 33 |
| 4. | (50-60) Years | 35 |
| 5. | (60-70) Years | 37 |

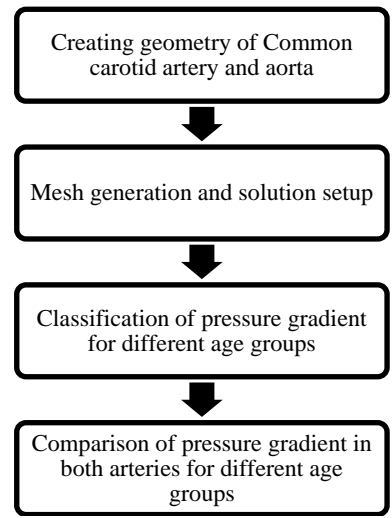

**Fig. 3: Practical block diagram**

Pressure gradient is calculated in left1 wall artery, middle1 wall artery and middle2 wall artery of CCA and ascending aorta of aorta. Here pressure gradient of CCA is compare with ascending aorta because in aorta aortic arch divide into three arteries i.e.; brachiocephalic artery, carotid artery and left subclavian artery. The oxygenated fresh blood from the heart via left ventricle enters the ascending aorta and then it is directed toward the aortic arch and comes out through descending aorta. From aortic arch blood is directed to CCA which direct blood to face, brain. Also the value pressure gradient of both the arteries will show which artery is more prone to arterial diseases.

In the current study, carreau model [4, 14] is used to model blood viscosity ($\mu$) with strain rate, $\gamma$ and is given by Eq. 2 shown below:

$$\mu = \mu_\infty + (\mu_o - \mu_\infty)[1 + (\lambda\gamma)^2]^{(n-1)/2} \ldots\ldots\ldots (2)$$

Here time constant ($\lambda$) is 3.313 s, n is 0.3568, zero strain viscosity (i.e. resting viscosity), ($\mu_0$) is 0.56 P and infinite strain viscosity ($\mu\infty$) is 0.0345 P. It's is given in Poise, P (where 1 P = 0.1 Ns/$m^2$ ). The blood density is considered as constant value: 1060 kg/m^3 and blood flow is pulsatile and cyclic. Also since blood is non-Newtonian fluid therefore the coefficient of viscosity of blood is function of velocity gradient.

## 3. RESULTS

The Pressure gradient of CCA according to different age group on left1, middle1 and middle2 wall artery for female group is shown in Fig. 4, 5 and 6.

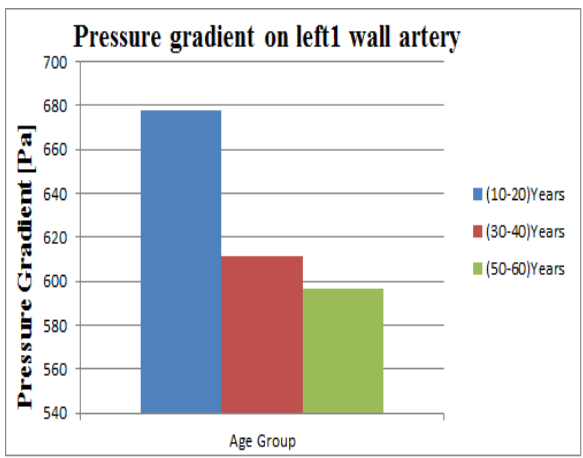

**Fig. 4: Pressure gradient on left1 wall artery of CCA of female group**

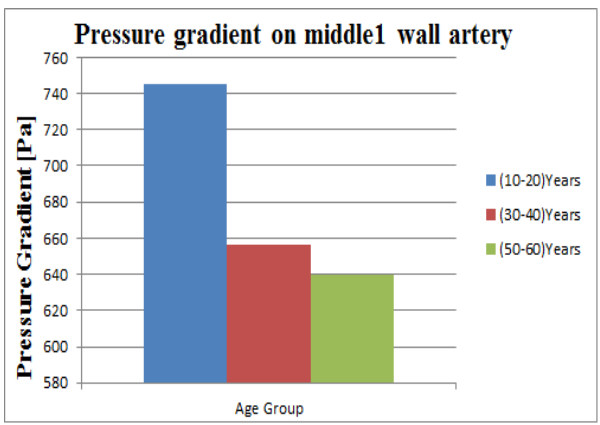

**Fig. 5: Pressure gradient on middle1 wallartery of CCA of female group**

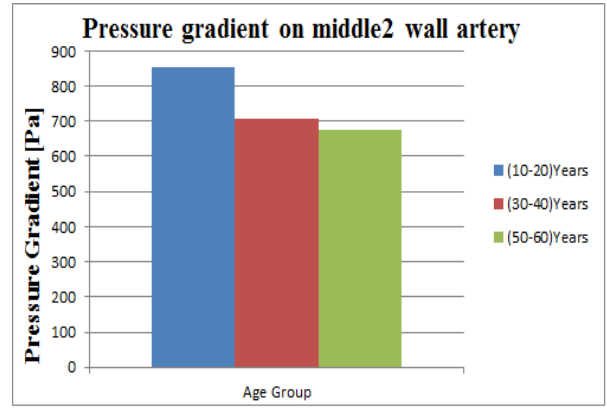

**Fig. 6: Pressure gradient on middle2 wall artery of CCA of female group**

Similarly the Pressure gradient of CCA according to different age group on left1, middle1 and middle2 wall artery for male group is shown in Fig. 7, 8 and 9.

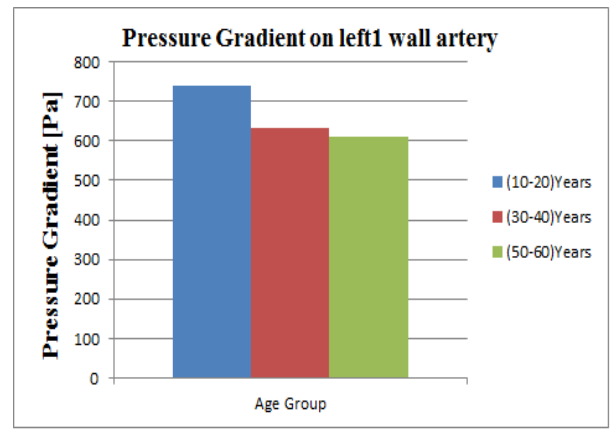

**Fig. 7: Pressure gradient on left1 wall artery of CCA of male group**

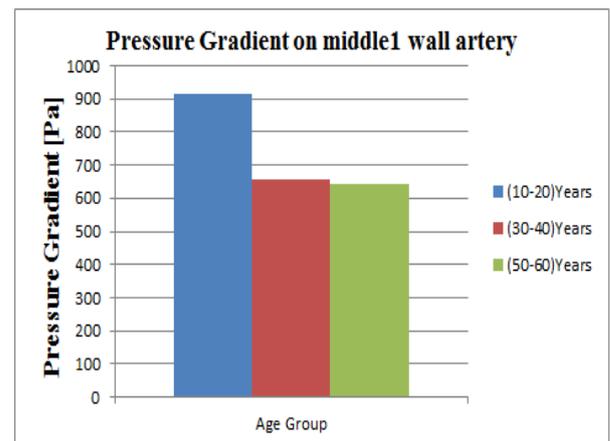

**Fig. 8: Pressure gradient on middle2 wall artery of CCA of male group**

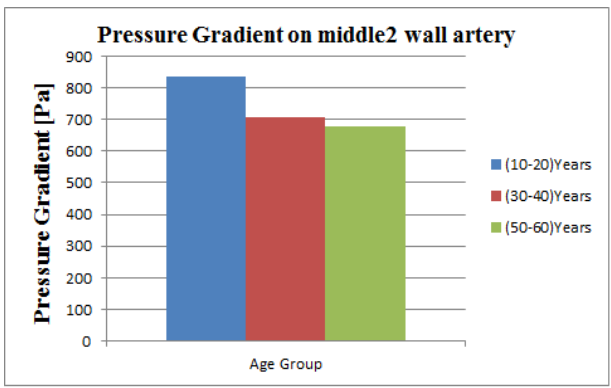

**Fig. 9: Pressure gradient on middle2 wall artery of CCA of male group**

It is found that with increase in diameter of CCA with advancing age pressure gradient decreases. This is due to decrease in driving force of flow due to increase in diameter of artery.

### 3.1 Pressure Gradient in CCA on the Basis of Gender

Pressure gradient of left1, middle1 and middle2 wall artery of CCA is classified on the basis of gender and is shown in Fig. 10, 11 and 12.

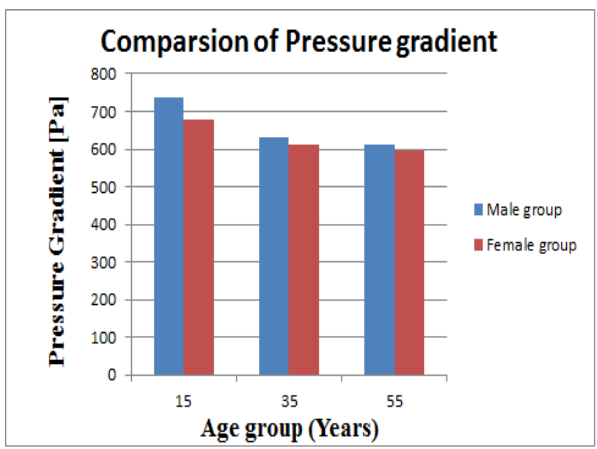

**Fig. 10: Comparison of pressure gradient of Carotid artery on left1 wall artery the basis of gender**

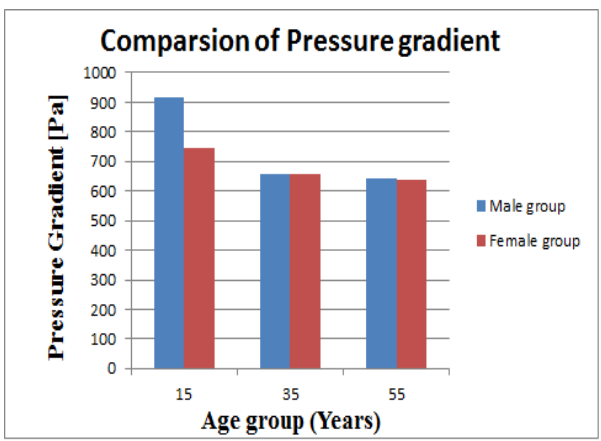

**Fig. 11: Comparison of pressure gradient of Carotid artery on middle1 wall artery the basis of gender**

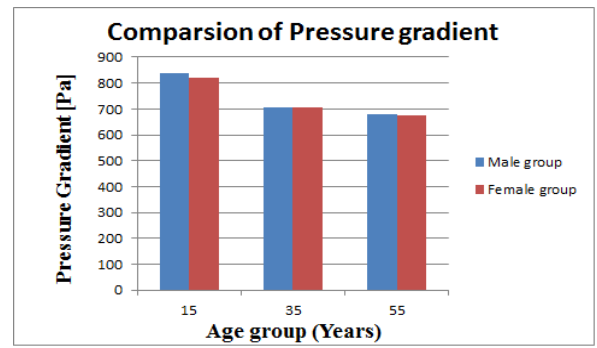

**Fig. 12: Comparison of pressure gradient of Carotid artery on middle2 wall artery the basis of gender**

It has been found that there is significant difference between pressure gradient of both male and female group. This difference is because of different body size, weight and height of both the gender.

### 3.2 Pressure Gradient in Aorta on the Basis of Age

The pressure gradient of ascending aorta according to different age group on ascending aorta is shown in Fig. 13.

It is found that with increase in diameter of ascending aorta with advancing age pressure gradient decreases. This is due to decrease in driving force of flow due to increase in diameter of artery.

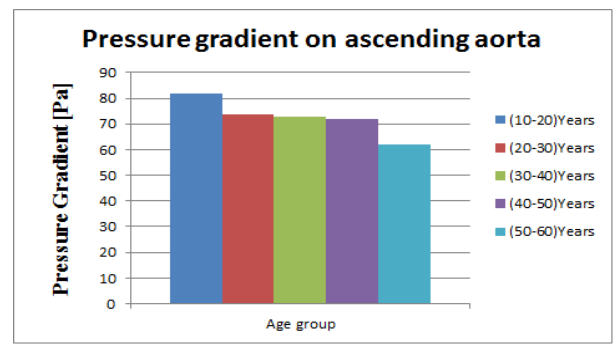

**Fig. 13: Pressure gradient of ascending aorta on the basis of age**

### 3.3 Comparison of Pressure Gradient in Both Arteries on the Basis of Age

Now the pressure gradient of both arteries for different age group is compared as shown in Fig. 14.

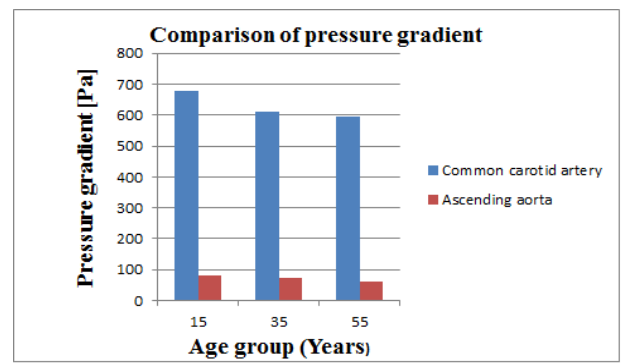

**Fig.14: Comparison of pressure gradient of both the arteries on the basis of age**

## 3.4 Comparison of Pressure Gradient in Both Arteries on the Basis of Gender

Also the pressure gradient of both arteries on the basis of gender is compared as shown in Fig. 15.

The graphs of comparison of pressure gradient in both the arteries on the basis of age and gender show that common carotid artery has greater value of pressure gradient than aorta in both the cases of age and gender.

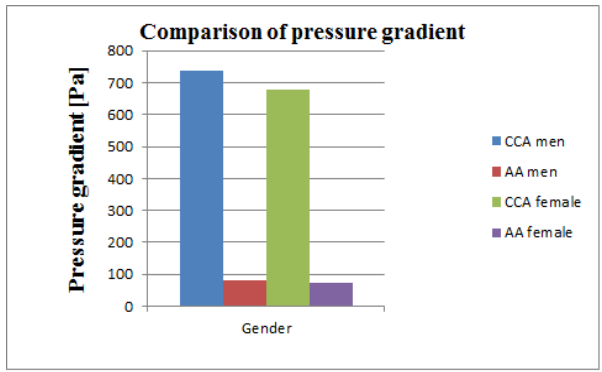

Fig.15: Comparison of pressure gradient of both the arteries on the basis of gender

## 4. CONCLUSION

The current study implies that the pressure gradient decline with advancing age in both cases. The decline in pressure gradient is due to decline in driving force of blood flow. But when pressure gradient of common carotid artery is compared with ascending aorta than its value is more in common carotid artery than aorta. So as the value of pressure gradient of ascending aorta is less as compared to common carotid artery therefore it has more risk of formation of plaque and hence more chance of having cardiovascular diseases as compared to common carotid artery. This study presents fast and economical method for modelling flow behaviour within arteries. Furthermore pressure gradient can be calculated in left subclavian artery as it supply blood to left arm and pain in left arm is one of the symptom of heart attack.